\documentclass[12pt]{iopart}
\usepackage{iopams} 
\usepackage{latexsym}
\usepackage{amsfonts}
\usepackage{amssymb}
\usepackage{color}
\usepackage[english]{babel}
\usepackage{setstack}
\usepackage{graphics}

\newtheorem{theorem}{Theorem}

\newtheorem{corollary}[theorem]{Corollary}
\newtheorem{definition}[theorem]{Definition}
\newtheorem{proposition}[theorem]{Proposition}
\newtheorem{remark}[theorem]{Remark}
\newtheorem{example}{Example}
\newenvironment{proof}[1][Proof]{\textbf{#1.} }{\ \rule{0.5em}{0.5em}
\vspace{0.3cm} }

\newcommand{\hi}{\mathcal{H}} 
\newcommand{\lh}{\mathcal{L(H)}} 
\newcommand{\sh}{\mathcal{S(H)}} 

\newcommand{\ip}[2]{\left\langle\,#1\,|\,#2\,\right\rangle} 

\newcommand{\kb}[2]{|#1\,\rangle\langle\,#2|} 
\newcommand{\no}[1]{\left\|#1\right\|} 
\newcommand{\com}{com} 
\newcommand{\ran}{{\rm ran}} 

\newcommand{\F}{\mathcal{F}}

\newcommand{\R}{\mathbb R} 
\newcommand{\RR}{\mathbb R^2}
\newcommand{\Z}{\mathbb Z} 

\newcommand{\A}{\mathcal{A}}
\newcommand{\bor}[1]{\mathcal{B}({#1})}
\newcommand{\borel}{\mathcal B(\R)}
\newcommand{\brr}{\mathcal B(\R^2)}
\newcommand{\brn}{\mathcal B(\R^n)}
\newcommand{\prob}{M_1^+(\R)}

\newcommand{\me}{m}

\begin{document}

\title[Coexistence of position and momentum]{On the
  coexistence of position and momentum observables}

\author{Claudio Carmeli\dag,\ Teiko Heinonen\ddag\  and Alessandro Toigo\dag}

\address{\dag\ Dipartimento di Fisica, Universit\`a di
  Genova, and I.N.F.N., Sezione di Genova, Via Dodecaneso 33, 16146
  Genova, Italy }

\address{\ddag\ Department of Physics, University of Turku, 
FIN-20014 Turku, Finland }

\eads{{\mailto{carmeli@ge.infn.it}},
  {\mailto{teiko.heinonen@utu.fi}}, {\mailto{toigo@ge.infn.it}}}

\begin{abstract}
We investigate the problem of coexistence of position and momentum observables.
We characterize those pairs of position and momentum observables which
have a joint observable.
\end{abstract}


\section{Introduction}\label{Introduction}

The problem of joint measurability of position and momentum
observables in quantum mechanics has a long history and different
viewpoints have been presented (see e.g. \cite{BL84}). According to
a common view sharp position and momentum observables 
are complementary quantities and therefore are not jointly
measurable. This is also illustrated, for instance, by the fact that
the Wigner distribution is not a probability distribution.
Advent of positive operator measures to quantum mechanics has made
further mathematically sound development possible. In this framework
certain observables, which are interpreted as unsharp position and
momentum observables, have joint measurements \cite{QTOS},
\cite{QDET}, \cite{PSAQT}.   
    
Although a collection of important results have been obtained, it
seems that the fundamental problem of joint measurability of position
and momentum observables has not yet been solved in its full
generality. Coexistent position and momentum observables (in the sense
of G. Ludwig \cite{FQMI}) have not been characterized so far.     

Our analysis of this problem proceeds in the following way. 
In Section \ref{Concepts} we fix the notations and recall some
concepts which are essential for our investigation. In Section
\ref{Position} position and momentum observables are defined through
their behaviour under the appropriate symmetry transformations. In
Section \ref{Joint} we follow a recent work of R. Werner
\cite{Werner04} to characterize those pairs of position and momentum
observables which are functionally coexistent and can thus be measured
jointly. Also some properties of joint observables are
investigated. In Section \ref{Coexistence} we present a few
observations on the general problem of coexistence of position and
momentum observables.

\section{Coexistence and joint observables}\label{Concepts} 

Let $\hi$ be a complex separable Hilbert space and $\lh$ the
set of bounded linear operators on $\hi$. The null operator and the identity
operator are denoted by $O$ and $I$, respectively. Let $\Omega$ be a
(nonempty) set and $\A$ a $\sigma$-algebra of subsets of $\Omega$. A
set function $E:\A\to\lh$ is an {\em operator measure} if it is
$\sigma$-additive (with respect to the weak operator topology). If
$E(X)\geq O$ for all $X\in\A$, we say that $E$ is \emph{positive}, and
$E$ is \emph{normalized} if $E(\Omega)=I$. The range of an
operator measure $E$ is denoted by $\ran (E)$, that is,  
\begin{equation*}
\ran (E)=\{ E(X)\mid X\in\A\}.
\end{equation*}

In quantum mechanics \emph{observables} are represented as normalized
positive operator measures and \emph{states} as positive operators of
trace one. We denote by $\sh$ the set of states. For an observable
$E:\A\to\lh$ and a state $T\in\sh$, we let $p^E_T$ denote the
probability measure on $\Omega$, defined by 
\begin{equation*}
p^E_T(X)=\tr[TE(X)],\quad X\in\A.
\end{equation*}
This is the probability distribution of the measurement outcomes when the
system is in the state $T$ and the observable $E$ is measured. 
If the range of $E$ contains only projections, then $E$
is called a \emph{sharp observable}. For more about observables as
normalized positive operator measures, the reader may refer to the
monographs \cite{PSAQT}, \cite{OQP}, \cite{QTCM}, and \cite{SSQT}. 

The notions of coexistence, functional coexistence and joint
observables are essential when the joint measurability of quantum
observables is analyzed. We next shortly recall the definitions of
these concepts. For further details we refer to a convenient survey
\cite{Lahti03} and references given therein. 

\begin{definition}
Let $(\Omega_i,\A_i)$, $i=1,2$, be measurable spaces and let 
$E_i:\A_i\to\lh$ be observables.
\begin{itemize}
\item[(i)] $E_1$ and $E_2$ are \emph{coexistent} if there is a
  measurable space $(\Omega,\A)$ and an observable $G:\A\to\lh$ such that 
\begin{equation*}
\ran(E_1)\cup\ran(E_2)\subseteq\ran(G).
\end{equation*}
\item[(ii)] $E_1$ and $E_2$ are \emph{functionally coexistent} if
  there is a measurable space $(\Omega,\A)$, an observable
  $G:\A\to\lh$ and measurable functions $f_1:\Omega\to\Omega_1$,
  $f_2:\Omega\to\Omega_2$, such that for any $X\in\A_1,Y\in\A_2$,
\begin{equation*}
E_1(X)=G(f^{-1}_1(X)),\quad E_2(Y)=G(f^{-1}_2(Y)).
\end{equation*}
\end{itemize} 
\end{definition} 

Functionally coexistent observables are coexistent, but it is an
open question if the reverse holds. 

We now confine our discussion to observables on $\R$. We denote by
$\brn$ the Borel $\sigma$-algebra of $\R^n$.

\begin{definition}
Let $E_1,E_2:\borel\to\lh$ be observables. An observable $G:\brr\to\lh$ is
their \emph{joint observable} if for all $X,Y\in\borel$, 
\begin{eqnarray*}
E_1(X) &=& G(X\times\R), \\
E_2(Y) &=& G(\R\times Y).
\end{eqnarray*}
In this case $E_1$ and $E_2$ are the \emph{margins} of $G$.
\end{definition}

For observables $E_1$ and $E_2$ defined on $\borel$ the existence of a
joint observable is equivalent to their functional
coexistence. These conditions are also equivalent to the \emph{joint
  measurability} of $E_1$ and $E_2$ in the sense of the quantum measurement
theory (see \cite{Lahti03}, Section 7).  

The \emph{commutation domain} of observables
$E_1$ and $E_2$, denoted by  $\com(E_1,E_2)$, is the closed subspace of
$\hi$ defined as  
\begin{equation*}
\fl \com(E_1,E_2)=\{\psi\in\hi\mid E_1(X)E_2(Y)\psi-E_2(Y)E_1(X)\psi=0\ \forall
X,Y\in\borel\}.
\end{equation*}
If $E_1$ and $E_2$ are sharp observables, then $E_1$ and $E_2$ are
coexistent only if they are functionally coexistent and this is the
case exactly when $\com(E_1,E_2)=\hi$. In general, for two observables
$E_1$ and $E_2$ the condition $\com(E_1,E_2)=\hi$ is sufficient but
not necessary for the functional coexistence of $E_1$ and $E_2$.   

In conclusion, given a pair of observables one may
pose the questions of their commutativity, functional coexistence, and
coexistence, in the order of increasing generality. 

\section{Position and momentum observables}\label{Position}

Let us shortly recall the standard description of a spin-0 particle in the
one-dimensional space $\R$. Fix $\hi=L^2(\R)$ and let $U$ and $V$ be
the one-parameter unitary representations on $\hi$, acting on
$\psi\in\hi$ as
\begin{eqnarray*}
\left[U(q)\psi\right](x) &=& \psi(x-q), \\
\left[V(p)\psi\right](x) &=& e^{ipx}\psi(x).
\end{eqnarray*}
Representations $U$ and $V$ correspond to space translations and velocity
boosts. They can be combined to form the following irreducible
projective representation $W$ of $\RR$, 
\begin{equation}\label{multiplier}
W(q,p)=e^{iqp/2}U(q)V(p).
\end{equation} 

Let $P$ and $Q$ be the selfadjoint operators generating $U$ and $V$,
that is, $U(q)=e^{-iqP}$ and $V(p)=e^{ipQ}$ for every $q,p\in\R$. We
denote by $\Pi_P$ and $\Pi_Q$ the sharp observables corresponding to the
operators $P$ and $Q$. For any $X\in\borel$ and $\psi\in\hi$ we then have
\begin{eqnarray}
 \Pi_Q(X)\psi & = & \chi_X\psi,  \nonumber\\
\Pi_P(X) & = & \F^{-1}\Pi_Q(X)\F, \label{P}
\end{eqnarray}
where $\chi_X$ is the characteristic function of $X$ and
$\F:\hi\to\hi$ is the Fourier-Plancherel operator. Sharp observables
$\Pi_Q$ and $\Pi_P$ correspond to position and momentum measurements
of absolute precision. We call them \emph{the canonical position
  observable} and \emph{the canonical momentum observable}, respectively.  

We take the symmetry properties of $\Pi_Q$ and $\Pi_P$ as the defining
properties of generic position and momentum observables. An observable
$E:\borel\to\lh$ is a \emph{position observable} if, for all $q,p\in\R$
and $X\in\borel$, 
\begin{eqnarray}
U(q)E(X)U(q)^* &=& E(X+q),\label{covE}\\
V(p)E(X)V(p)^* &=& E(X).\label{invE}
\end{eqnarray}
This means that a position observable is defined as a translation
covariant and velocity boost invariant observable. In our previous article
\cite{CHT04} we have shown that these conditions are satisfied exactly
when there is a probability measure $\rho:\borel\to[0,1]$ such that    
\begin{equation}\label{EX}  
E(X)=E_{\rho}(X):=\int \rho(X-q)\ d\Pi_Q(q),\quad X\in\borel, 
\end{equation} 
where $X-q=\{x-q\mid x\in X\}$. A position observable $E_{\rho}$ can be
interpreted as a \emph{fuzzy version} of the canonical position
observable $\Pi_Q$, unsharpness being characterized by the probability
measure $\rho$ (see e.g. \cite{AP77b}, \cite{AD76},
\cite{Busch85}). We call  $E_{\rho}$ a \emph{fuzzy position
  observable} if $E_{\rho}$ is not a sharp observable.   

We denote by $M(\R)$ the set of complex measures on $\R$ and $\prob$
is the subset of probability measures. For any $\lambda\in M(\R)$, 
$\widehat{\lambda}$ denotes the Fourier-Stieltjes transform of $\lambda$. 

\begin{proposition}\label{inj}
Let $\rho_1,\rho_2\in\prob$, $\rho _{1}\neq \rho _{2}$. Then
$E_{\rho _{1}}\neq E_{\rho _{2}}$. 
\end{proposition}

\begin{proof}
For $\psi \in \mathcal{H}$, we define the real measure $\lambda_{\psi}$ by
$$
\lambda_{\psi }\left(
X\right) =\ip{\psi}{\left( E_{\rho _{1}}\left( X\right) -E_{\rho
_{2}}\left( X\right) \right) \psi} =\mu_{\psi }\ast \left(
\rho _{1}-\rho _{2}\right) \left( X\right), 
$$ 
where $\ast$ is the convolution and $d\mu _{\psi
}\left( x\right) =\left| \psi \left( x\right) \right| ^{2}dx$. Taking
the Fourier transform we get 
\begin{equation*}
\widehat{\lambda}_{\psi }=\widehat{\mu}_{\psi }\cdot \left( 
\widehat{\rho_1} -\widehat{\rho_2}\right),
\end{equation*}
where $\widehat{\lambda}_{\psi }$, $\widehat{\mu}_{\psi }$, $\widehat{\rho_1}$
and $\widehat{\rho_2}$ are
continuous functions. By injectivity of the Fourier-Stieltjes transform
we have $\widehat{\rho_1}\neq \widehat{\rho_2}$. Thus, choosing $\psi$
such that $\widehat{\left| \psi \right|^{2}}(p)\neq 0$
for every $p\in\R$, we have $\widehat{\lambda}_{\psi }\neq 0$. This
means that $\lambda_{\psi}\neq 0$ and hence, $E_{\rho _{1}}\neq E_{\rho _{2}}$.
\end{proof}

By Proposition \ref{inj} there is one-to-one correspondence between
the set of position observables and  $\prob$. A position observable
$E_{\rho}$ is a sharp observable if and only if $\rho=\delta_{x}$ for
some $x\in\R$, where $\delta_x$ is the Dirac measure concentrated at $x$,
\cite{CHT04}. Since the Dirac measures are the extreme elements of the
convex set $\prob$, the sharp position observables are the extreme
elements of the set of position observables. The canonical position
observable $\Pi_Q$ corresponds to the Dirac measure $\delta_0$. 

In an analogous way, a momentum observable is defined as a velocity boost
covariant and translation invariant observable. Thus, an observable
$F:\borel\to\lh$ is a \emph{momentum observable} if, for all $q,p\in\R$
and $Y\in\borel$, 
\begin{eqnarray}
V(p)F(Y)V(p)^* &=& F(Y+p),\label{covF} \\
U(q)F(Y)U(q)^* &=& F(Y).\label{invF}
\end{eqnarray}
Since an observable $E$ is a position observable if and
only if $\F^{-1}E\F$ is a momentum observable, the previous discussion on
position observables is easily converted to the case of momentum
observables. In particular, an observable $F$ satisfies conditions
(\ref{covF}) and (\ref{invF}) if and only if there is a probability
measure $\nu:\borel\to [0,1]$ such that $F=F_{\nu}$, where 
\begin{equation}\label{FY}
F_{\nu}(Y):=\int \nu(Y-p)\ d\Pi_P(p),\quad Y\in\borel.
\end{equation}

For completeness we give a proof of the following known fact
\cite{BL89}, which will be needed later.  

\begin{proposition}\label{noncommu}
A position observable $E_{\rho}$ and a momentum observable $F_{\nu}$
are totally noncommutative, that is, $\com(E_{\rho},F_{\nu})=\{0\}$.
\end{proposition} 
 
\begin{proof}
It is shown in \cite{BSS87} and \cite{Ylinen89} that for functions $f,g\in
L^{\infty}(\R)$ the equation 
\begin{equation*}
f(Q)g(P)-g(P)f(Q)=O
\end{equation*}
holds if and only if one of the following is satisfied: (i) either
$f(Q)$ or $g(P)$ is a multiple of the identity operator, (ii) $f$ and
$g$ are both periodic with minimal periods $a,b$ satisfying
$2\pi/ab\in\Z\smallsetminus \{0\}$. 

Let $X\subset\R$ be a bounded interval. Then the operators
$E_{\rho}(X)$ and $F_{\nu}(X)$ are not multiples of the identity
operator. Indeed, let us assume, in contrary, that $E_{\rho}(X)=cI$
for some $c\in\R$. Denote $a=2 |X|$, where $|X|$ is the lenght of
$X$. Then the sets $X+na$, $n\in\Z$, 
are pairwisely disjoint and  
\begin{eqnarray*}
I &\geq& E_{\rho}(\cup_{n\in\Z}(X+na))=\sum_{n=-\infty}^{\infty}
 E_{\rho}(X+na)\\
 &=& \sum_{n=-\infty}^{\infty} U(na)E_{\rho}(X)U(na)^* =
 \sum_{n=-\infty}^{\infty}  cI.
\end{eqnarray*}
This means that $c=0$. However, since $|X|>0$, we have
$E_{\rho}(X)\neq O$ (see e.g. \cite{HLPPY03}). Thus,
\begin{equation*}
O\neq E_{\rho}(X)=cI=O
\end{equation*}
and $E_{\rho}(X)$ is not a multiple of the
identity operator. Moreover, since $\rho(\R)=1$, the function
$q\mapsto\rho(X-q)$ is not periodic. We conclude that, by the above
mentioned result, the operators $E_{\rho}(X)$ and $F_{\nu}(X)$ do not 
commute and hence, $\com(E_{\rho},F_{\nu})\neq\hi$. 

Assume then that there exists $\psi\neq 0$,
$\psi\in\com(E_{\rho},F_{\nu})$. Using the symmetry properties
(\ref{covE}), (\ref{invE}), (\ref{covF}) and (\ref{invF}), a short
calculation shows that for any $q,p\in\R$,
$U(q)V(p)\psi\in\com(E_{\rho},F_{\nu})$. This implies that
$\com(E_{\rho},F_{\nu})$ is invariant under the irreducible projective
representation $W$ defined in (\ref{multiplier}). As
$\com(E_{\rho},F_{\nu})$ is a closed subspace of $\hi$, it follows that either
$\com(E_{\rho},F_{\nu})=\{0\}$ or $\com(E_{\rho},F_{\nu})=\hi$. Since
the latter possibility is ruled out, this completes the proof.     
\end{proof}

\section{Joint observables of position and momentum
observables}\label{Joint}

Looking at the symmetry conditions (\ref{covE}), (\ref{invE}),
(\ref{covF}) and (\ref{invF}), and equation (\ref{multiplier}), it is 
clear that an observable $G:\brr\to\lh$ has a position observable and
a momentum observable as its margins if and only if, for all $q,p\in\R$
and
$X,Y\in\borel$, the following conditions hold: 
\begin{eqnarray}
W(q,p)G(X\times\R)W(q,p)^* &=& G(X\times\R+(q,p)), \label{WW1}\\
W(q,p)G(\R\times Y)W(q,p)^* &=& G(\R\times Y+(q,p)).\label{WW2}
\end{eqnarray}

\begin{definition}\label{cpso}
An observable $G:\bor{\R^2}\to\lh$ is a covariant phase space
observable if for all $q,p\in\R$ and $Z\in\bor{\R^2}$,
\begin{equation}\label{W}
W(q,p)G(Z)W(q,p)^*=G(Z+(q,p)).
\end{equation}
\end{definition}

It is trivial that (\ref{W}) implies (\ref{WW1}) and (\ref{WW2}) and,
hence, a covariant phase space observable is a joint observable of some
position and momentum observables. To our knowledge, it is an
open question whether (\ref{WW1}) and (\ref{WW2}) imply (\ref{W}).

For any $T\in\sh$, we define an observable $G_T:\brr\to\lh$ by 
\begin{equation}
G_T(Z)=\frac{1}{2\pi}\int_Z W(q,p)TW(q,p)^*\ dqdp,\quad Z\in\brr.
\end{equation}
The observable $G_T$ is a covariant phase space observable. Moreover,
if $G$ is a covariant phase space observable, then $G=G_T$ for some
state $T\in\sh$, \cite{PSAQT}, \cite{CDT04},\cite{Werner84}. 

\begin{proposition}
Let $T_1,T_2\in\sh$, $T_1\neq T_2$. Then $G_{T_1}\neq G_{T_2}$.
\end{proposition}

\begin{proof}
Let us first note that for any $T,S\in\sh$ and $Z\in\brr$,
\begin{eqnarray*}
p^{G_T}_S(Z) &=& \frac{1}{2\pi}\int_Z \tr[S W(q,p)TW(q,p)^*]\ dqdp \\
&=& \frac{1}{2\pi}\int_Z \tr[TW(q,p)^* S W(q,p)]\ dqdp \\
&=& \frac{1}{2\pi}\int_{-Z} \tr[TW(q,p) S W(q,p)^*]\ dqdp \\
&=& p^{G_S}_T(-Z).
\end{eqnarray*}
\noindent
Let $T_1,T_2\in\sh$ and assume that $G_{T_1}=G_{T_2}$. This means that for
any $S\in\sh$,
\begin{equation}\label{GT}
p^{G_{T_1}}_S=p^{G_{T_2}}_S,
\end{equation}
which is, by the previous observation, equivalent to
\begin{equation}\label{GS}
p^{G_S}_{T_1}=p^{G_S}_{T_2}.
\end{equation}

Let $S$ be a state such that $G_S$ is an informationally
complete observable (see \cite{AP77a}). Then (\ref{GS}) implies that
$T_1=T_2$.   
\end{proof}

Let $G_T$ be a covariant phase space observable and let $\sum_i \lambda_i
\kb{\varphi_i}{\varphi_i}$ be the spectral decomposition of the state $T$. 
The margins of $G_T$ are a position observable $E_{\rho}$ and a
momentum observable $F_{\nu}$, where 
\begin{eqnarray}
&& d\rho(q) = e(q)dq,\quad e(q)=\sum_i\lambda_i
|\varphi_i(-q)|^2,\label{marginrho}\\ 
&& d\nu(p) = f(p)dp,\quad f(p)=\sum_i\lambda_i
|\widehat{\varphi_i}(-p)|^2.\label{marginnu} 
\end{eqnarray}  
The form of $\rho$ and $\nu$ in (\ref{marginrho}) and (\ref{marginnu})
implies that, in general, the margins $E_{\rho}$ and $F_{\nu}$ do not
determine $G_T$, that is, another covariant phase space observable
$G_{T'}$ may have the same margins. Indeed, the functions
$|\varphi(\cdot)|$ and $|\widehat{\varphi}(\cdot)|$ do not define the
vector $\varphi$ uniquely up to a phase factor. (This is also known as
the Pauli problem).
 
\begin{example}\rm\label{sopra}
Consider the functions 
\begin{equation}
\label{gauss1}
\varphi_{a,b}\left( q\right) =\left( \frac{2a}{\pi }\right)
^{1/4}e^{-\left(
a+ib\right) q^{2}},
\end{equation}
with $a,b\in \mathbb{R}$ and $a>0$. The Fourier transform of
$\varphi_{a,b}$ is
\begin{eqnarray}
\widehat{\varphi}_{a,b}\left( p\right)  &=&\left( \frac{a}{2\pi \left(
a^{2}+b^{2}\right) }\right) ^{1/4}\exp \left( -\frac{ap^{2}}{4\left(
a^{2}+b^{2}\right) }\right) \nonumber\\
\label{gauss2}
&&\times\exp \left( \frac{ibp^{2}}{4\left( a^{2}+b^{2}\right)
}-\frac{i}{2}\arctan 
\frac{b}{a}\right) .
\end{eqnarray}
For $b\neq 0$, we see that $T_1=\kb{\varphi_{a,b}}{\varphi_{a,b}}$ and
$T_2=\kb{\varphi_{a,-b}}{\varphi_{a,-b}}$ are different, but the margins
of $G_{T_1}$ and $G_{T_2}$ are the same position and momentum observables
$E_\rho$ and $F_\nu$, with
\begin{eqnarray*}
 d\rho(q) & = & \left( \frac{2a}{\pi }\right) ^{1/2}e^{-2a q^{2}} dq\\
d\nu(p) & = & \left( \frac{a}{2\pi \left(
a^{2}+b^{2}\right) }\right) ^{1/2}\exp \left( -\frac{ap^{2}}{2\left(
a^{2}+b^{2}\right) }\right) dp
\end{eqnarray*}  
\end{example}
\vspace{0.3cm}

As $\rho$ and $\nu$ in (\ref{marginrho}) and (\ref{marginnu}) arise
from the same state $T$, a multitude of
uncertainty relations can be derived for the observables $E_{\rho}$
and $F_{\nu}$. One of the most common uncertainty relation is in terms of
variances. Namely, let $Var(p)$ denote the variance of a probability
measure $p$, 
\begin{equation*}
Var(p)= \int \left( x - \int x\ dp(x) \right)^2\ dp(x). 
\end{equation*}
Then for any state $S$,
\begin{equation}\label{ur}
Var(p^{E_{\rho}}_S)\ Var(p^{F_{\nu}}_S)\geq 1.
\end{equation}
(See e.g. \cite{OQP}, Section III.2.4 or \cite{CRQM}, Section 5.4.)
The lower bound in (\ref{ur}) can be achieved only if 
\begin{equation}\label{min}
Var(\rho)\ Var(\nu)=\frac{1}{4},
\end{equation}
and it is well known that (\ref{min}) holds if and only if
$T=\kb{\varphi}{\varphi}$ and $\varphi$ is a Gaussian function of the form 
\begin{equation} \label{gauss3}
\varphi \left( q\right) =\left(2a/\pi \right) ^{1/4}e^{ibq}e^{-a\left(
  q-c\right) ^{2}},\quad  a>0,\ b,c \in\R.
\end{equation}
It is also easily verified that choosing $S=T$ the equality in
(\ref{ur}) is indeed obtained. 

\begin{proposition}\label{fcoex1}
Let $E_{\rho}$ be a position observable and $F_{\nu}$ a momentum
observable. If $E_{\rho}$ and $F_{\nu}$ have a joint observable, then
they also have a joint observable which is a covariant phase space
observable. 
\end{proposition}

The proof of Proposition \ref{fcoex1} is given in Appendix A.

\begin{corollary}\label{fcoex2}
A position observable $E_{\rho}$ and a momentum observable $F_{\nu}$
are functionally coexistent if and only if there  
is a state $T\in\sh$ such that $\rho$ and $\nu$ are given by
(\ref{marginrho}) and (\ref{marginnu}). In particular, the uncertainty
relation (\ref{ur}) is a necessary condition for functional
coexistence, and thus, for the joint measurability of $E_{\rho}$ and
$F_{\nu}$.  
\end{corollary}

\begin{remark}\rm\label{FP}
As the canonical position observable and the canonical momentum
observable are Fourier equivalent (see (\ref{P})), one may also want to
require this connection from a fuzzy 
position observable $E_{\rho}$ and a fuzzy momentum observable
$F_{\nu}$. This requirement, in general, simply leads to condition
$\rho=\nu$. Let us consider the case when $E_{\rho}$ and $F_{\nu}$ are
the margins of the covariant phase space observable $G_T$ generated by
a pure state $T=\kb{\varphi}{\varphi}$. Then $\rho=\nu$ exactly when
$\left| \varphi \right| =\left| \widehat{\varphi}\right| $.  To give an
example when this condition is satisfied, suppose $\varphi=\varphi_{a,b}$,
with $\varphi_{a,b}$ defined in equation (\ref{gauss1}). By equation
(\ref{gauss2}), the condition $\left| \varphi_{a,b}\right| =\left|
\widehat{\varphi}_{a,b}\right| $ is equivalent to
\begin{equation}\label{ab}
a^{2}+b^{2}=1/4.
\end{equation}
Thus, if the numbers $a$ and $b$ are chosen so that they satisfy
(\ref{ab}), the vector $\varphi_{a,b}$ defines Fourier equivalent
  position and momentum observables. 
\end{remark}

We end this section with an observation about a (lacking) localization
property of a joint observable of position and momentum observables. We
wish to emphasize that $G$ in Proposition \ref{Gprop} is not assumed to be
a covariant phase space observable.  

\begin{proposition}\label{Gprop}
Let $G$ be a joint observable of a position observable $E_{\rho}$ and
a momentum observable $F_{\nu}$ and let $Z\in\brr$ be a bounded set. Then 
\begin{itemize}
\item[(i)] $\no{G(Z)}\neq 1$;
\item[(ii)] there exists a number $k_Z<1$ such that for any $T\in\sh$,
  $p^G_T(Z)\leq k_Z$.  
\end{itemize}
\end{proposition}

\begin{proof}

(i) It follows from Proposition \ref{fcoex1} and the Paley-Wiener
  theorem that either $\rho$ or $\nu$ has an unbounded support. Let us
  assume that, for instance, $\rho$ has an unbounded support. 

Let $Z\in\brr$ be a bounded set. Then the closure $\bar{Z}$ is compact
and also the set 
\begin{equation*}
X:=\{x\in\R\mid \exists y\in\R: (x,y)\in\bar{Z}\}\subset\R
\end{equation*}
is compact.
Since 
\begin{equation}
\no{G(Z)}\leq\no{G(X\times\R}=\no{E_{\rho}(X)}
\end{equation}
and
\begin{equation}
\no{E_{\rho}(X)}={\rm ess\,sup}_{x\in\R}\rho(X-x)\leq {\rm
  sup}_{x\in\R}\rho(X-x), 
\end{equation}
it is enough to show that
\begin{equation}\label{sup1}
{\rm sup}_{x\in\R}\rho(X-x)<1.
\end{equation}
Let us suppose, in contrary, that 
\begin{equation}\label{sup2}
{\rm sup}_{x\in \R} \rho(X-x)=1
\end{equation}
This means that there exists a sequence $(x_n)_{n\geq 1}\subset\R$ such
that 
\begin{equation}\label{lim}
\lim_{n\to\infty}\rho(X-x_n)=1.
\end{equation}
Since $\rho(\R)=1$ and $X$ is a bounded set, the sequence $(x_n)_{n\geq1}$ 
is also bounded. It follows that $B:=\bigcup_{n=1}^{\infty}\ X-x_n$ is
a bounded set and by (\ref{lim}) 
we have $\rho(B)=1$. 
This is in contradiction with the assumption that $\rho$ has an
unbounded support. Hence, (\ref{sup2}) is false and (\ref{sup1}) follows.

(ii) From (i) it follows that
\begin{equation*}
1 > k_Z := \no{G(Z)}\,=\,\sup \{ \ip{\psi}{G(Z)\psi}\mid
\psi\in\hi,\no{\psi}=1\}.
\end{equation*}
Let $T\in\sh$ and  let $\sum_i \lambda_i\kb{\psi_i}{\psi_i}$ be
the spectral decomposition of $T$. Then
\begin{equation*}
p^G_T(Z) = \tr[TG(Z)] = \sum_i \lambda_i \ip{\psi_i}{G(Z)\psi_i} \leq k_Z.
\end{equation*}
\end{proof}

\section{Coexistence of position and momentum observables}\label{Coexistence}

Since coexistence is, a priori, a more general concept than functional
coexistence, we are still left with the problem of characterizing
coexistent pairs of position and momentum observables. In lack of a
general result we close our investigation with some observations on
this problem.

\begin{proposition}\label{projection}
Let $E_{\rho}$ be a position observable and $F_{\nu}$ a momentum
observable. If $\ran(E_{\rho})\cup\ran(F_{\nu})$ contains a nontrivial
projection (not equal to $O$ or $I$), then 
$E_{\rho}$ and $F_{\nu}$ are not coexistent.
\end{proposition} 

\begin{proof}
Let us assume, in contrary, that there exists an observable $G$ such
that $\ran(E_{\rho})\cup\ran(F_{\nu})\subseteq\ran(G)$. Suppose, for
instance, that $E_{\rho}(X)$ is a nontrivial projection. 
Then $E_{\rho}(X)$ commutes with all operators in the range
of $G$ (see e.g. \cite{LP01}). In particular, $E_{\rho}(X)$
commutes with all $F_{\nu}(Y)$, $Y\in\borel$. However, this is impossible
by the result proved in \cite{BSS87} and \cite{Ylinen89} (see the
beginning of the proof of Proposition \ref{noncommu}).  
\end{proof}

\begin{corollary}\label{convex}
Let $E_{\rho}$ be a position observable which is a convex combination
of two sharp position observables. Then $E_{\rho}$ is not coexistent with
any momentum observable $F_{\nu}$. 
\end{corollary}

\begin{proof}
Let $E_{\rho _{1}}$, $E_{\rho _{2}}$ be sharp position observables with
$\rho _{1}=\delta _{a}$, $\rho _{2}=\delta _{b}$, and $E_{\rho}=tE_{\rho
_{1}}+\left( 1-t\right) E_{\rho _{2}}$ for some $0\leq t\leq 1$. This
means that $\rho =t\delta _{a}+\left( 1-t\right)
\delta _{b}$. If $a=b$, or $t\in\{0,1\}$, then $E_{\rho}$ is a sharp
observable and the claim follows from Proposition
\ref{projection}. Let us then assume that $a<b$ and $0<t<1$. Take
$X=\bigcup_{n\in \mathbb{Z}}\left[ n\left( b-a\right),\left(
  n+1/2\right) \left( b-a\right) \right] $. For any $x\in\mathbb{R}$, 
we have
\begin{eqnarray*}
\rho \left( X-x\right)  &=& t\delta_x(X-a) + (1-t)\delta_x(X-b)\\ 
&=& t\sum_{n=-\infty}^{\infty}\delta _{x}\left( \left[ nb-\left(
n+1\right) a,\left( n+1/2\right) b-\left( n+3/2\right) a\right] \right)
\\
&&+\left( 1-t\right) \sum_{n=-\infty}^{\infty}\delta _{x}\left( \left[
  \left( n-1\right) b-na,\left( n-1/2\right) b-\left( n+1/2\right)
  a\right] \right)  \\ 
&=&\sum_{n=-\infty}^{\infty}\delta _{x}\left( \left[ nb-\left( n+1\right)
a,\left(
n+1/2\right) b-\left( n+3/2\right) a\right] \right)\\
&=& \delta_x(X-a) 
\end{eqnarray*}
It follows that $E_{\rho}\left( X\right)
=\Pi_{Q}\left(X-a\right)$. Since the projection
$\Pi_{Q}\left(X-a\right)$ is nontrivial, the claim follows from
Proposition \ref{projection}.  
\end{proof}

Evidently, Corollary \ref{convex} has also a dual statement with the
roles of position and momentum observables reversed.

\ack
The authors wish to thank Paul Busch, Gianni Cassinelli, Pekka Lahti
and Kari Ylinen for useful comments on this paper.

\appendix

\section*{Appendix. Proof of Proposition \ref{fcoex1}}\label{proof1}
\setcounter{section}{1}

In order to prove Proposition \ref{fcoex1} we need some general results
about means on topological spaces, and for convenience they are 
briefly reviewed. The following material is based on \cite{AHAI},
Chapter~IV, \S 17, and \cite{Werner04}.

Let $\Omega$ be a locally compact separable metric space with
a metric $d$. By $BC\left( \Omega\right) $ we denote the Banach space
of complex 
valued bounded continuous functions on $\Omega$, with the uniform norm
$\left\| f\right\|_{\infty }=\sup_{x\in \Omega}\left| f\left( x\right)
\right|$. The linear subspace of continuous 
functions with compact support is denoted by $C_{c}\left(
\Omega\right)$. Adding the index $^{r}$ we denote the 
subsets of real functions in $BC\left(\Omega\right) $ or in $C_{c}\left(
\Omega\right) $. With the index $^{+}$ we denote the subsets of
positive functions. 

\begin{definition}
A \emph{mean} on $\Omega$ is a linear functional 
\begin{equation*}
\me:BC\left( \Omega\right)\longrightarrow \mathbb{C}
\end{equation*}
such that: 
\begin{itemize}
\item[(i)] $\me \left( f\right) \geq 0$ if $f\in BC^{+}\left(
  \Omega\right)$;
\item[(ii)] $\me \left( 1\right)=1$.
\end{itemize}
For a mean $\me$ on $\Omega$ we denote 
\begin{equation*}
\me \left( \infty \right) =1-\sup \left\{ \me \left( f \right) \mid
f \in C_{c}^{+}\left( \Omega\right) ,\ f \leq 1\right\} .
\end{equation*}
\end{definition}

Let $\me $ be a mean on $\Omega$. By the Riesz representation theorem,
there exists a unique positive Borel measure $\me_{0}$ on $\Omega$ such
that 
\begin{equation*}
\me \left( f \right) =\int_{\Omega}f \left( x\right) d\me
_{0}\left( x\right) \quad \forall f \in C_{c}\left( \Omega\right) .
\end{equation*}
By the inner regularity of $\me_{0}$ we have 
\begin{equation*}
\me_{0}\left( \Omega\right) =\sup \left\{ \me \left( f \right) \mid
f \in C_{c}^{+}\left( \Omega\right) ,\ f \leq 1\right\} =1-\me \left(
\infty \right) \leq 1.
\end{equation*}
In particular, any function in $BC\left( \Omega\right)$ is integrable with
respect to $m_0$. For any $f\in BC\left( \Omega\right)$, we use the
abbreviated notation  
\begin{equation*}
\me _{0}\left( f\right) :=\int_{\Omega}f\left( x\right) d\me _{0}\left(
x\right) .
\end{equation*}

\begin{proposition}\label{minfty}
If $\me \left( \infty \right) =0$, then
\begin{equation*}
\me \left( f\right) =\me _{0}\left( f\right) \quad \forall f\in BC\left(
\Omega\right) .
\end{equation*}
\end{proposition}

\begin{proof}
We fix a point $x_{0}\in \Omega$. For all $R>0$ we define 
\begin{equation*}
g_{R}\left( x\right) =\left\{ 
\begin{array}{ccc}
1 & \mbox{if} & d\left( x_{0},x\right) \leq R/2, \\ 
3/2-d\left( x_{0},x\right) /R & \mbox{if} & R/2<d\left( x_{0},x\right)
\leq
3R/2, \\ 
0 & \mbox{if} & d\left( x_{0},x\right) >3R/2.
\end{array}
\right. 
\end{equation*}
Then $g_{R}\in C_{c}^{+}\left( \Omega\right) $ and $g_{R}\leq
1$. Moreover, for any $f \in C_{c}^{+}\left( \Omega\right) $ such that $f
\leq 1$ there exists $R>0$ such that $f \leq g_{R}$,
and hence 
\begin{equation*}
1=\sup \left\{ \me \left(f  \right) \mid f \in C_{c}^{+}\left(
\Omega\right) ,\ f \leq 1\right\} =\lim_{R\rightarrow \infty }\me \left(
g_{R}\right) .
\end{equation*}

Let $f\in BC^{+}\left( \Omega\right)$ and $R>0$. Since $g_{R}f\in
C_{c}\left(\Omega\right)$, we have 
\begin{equation} \label{uno}
\me \left( f\right) =\me _{0}\left( g_{R}f\right) +\me \left( \left(
1-g_{R}\right) f\right).
\end{equation}
We have $0\leq g_{R}f\leq f$, $f$ is $\me_0$-integrable and
$\lim_{R\rightarrow \infty }g_{R}\left( x\right) 
f\left( x\right) =f\left( x\right)$ for all $x\in \Omega$. Therefore,
by the dominated convergence theorem we have 
\begin{equation*}
\lim_{R\rightarrow \infty }\int_{\Omega} g_{R}\left( x\right) f\left(
x\right) d \me _{0}\left( x\right) =\int_{\Omega}f\left( x\right) d
\me_{0}\left( x\right) .
\end{equation*}
For the other term in the sum (\ref{uno}), we have 
\begin{equation*}
\me \left( \left( 1-g_{R}\right) f\right) \leq \left\| f\right\|
_{\infty }\me \left( 1-g_{R}\right) \underset{R\rightarrow \infty }{
\longrightarrow }\left\| f\right\| _{\infty }\me \left( \infty \right) =0.
\end{equation*}
Taking the limit $R\rightarrow \infty $ in (\ref{uno}) we then get 
\begin{equation*}
\me \left( f\right) =\me _{0}\left( f\right) .
\end{equation*}

If $f\in BC\left( \Omega\right) $, we write $f=f_{1}+if_{2}$ with
$f_{1},f_{2}\in 
BC^{r}\left( \Omega\right) $, and $f_{i}=f_{i}^{+}-f_{i}^{-}$ with
$f_{i}^{\pm }= 
\frac{1}{2}\left( \left| f_{i}\right| \pm f_{i}\right) \in BC^{+}\left(
\Omega\right) $, and we use the previous result to obtain the conclusion.
\end{proof}

Let $i\in \left\{ 1,2\right\}$. For each $f\in BC\left( \Omega\right)$ we
define 
\begin{equation*}
\widetilde{f}_{i}\left( x_{1},x_{2}\right) := f\left(
x_{i}\right)\quad  \forall x_{1},x_{2}\in \Omega.
\end{equation*}
Clearly, $\widetilde{f}_{i}\in BC\left( \Omega\times \Omega\right)
$. For a mean $\me : 
BC\left( \Omega\times \Omega\right) \longrightarrow \mathbb{C}$, we
then define  
\begin{equation*}
\me _{i}\left( f\right) :=\me \left( \widetilde{f}_{i}\right) \quad
\forall
f\in BC\left( \Omega\right) .
\end{equation*}
The linear functional $\me_{i}:$ $BC\left( \Omega\right)
\longrightarrow \mathbb{C}$ is a mean on $\Omega$, which we call the
$i$th {\em margin} of $\me$. 

\begin{proposition}\label{mminfty}
Let $\me $ be a mean on $\Omega\times \Omega$. If $\me _{1}\left(
\infty \right) =\me_{2}\left( \infty \right) =0$, then $\me \left(
\infty \right) =0$. 
\end{proposition}

\begin{proof}
For all $R>0$, we define the function $g_{R}\in C_{c}\left( \Omega\right)$
as in the proof of Proposition~\ref{minfty}. We set 
\begin{equation*}
h_{R}\left( x_{1},x_{2}\right) =g_{R}\left(
x_{1}\right)g_{R}\left( x_{2}\right) .
\end{equation*}
Clearly, $h_{R}\in C_{c}^{+}\left( \Omega\times \Omega\right) $, and, if
$h \in
C_{c}^{+}\left( \Omega\times \Omega\right)$ and $h \leq 1$, there exists
$R>0$
such that $h \leq h_{R}$. Since 
\begin{eqnarray*}
1-h_{R}\left( x_{1},x_{2}\right)  &=&\left( 1-g_{R}\left(
x_{1}\right) \right) + g_{R}\left( x_{1}\right) \left( 1-
g_{R}\left( x_{2}\right) \right)  \\
&\leq &\left( 1-g_{R}\left( x_{1}\right) \right) +\left( 1-g_{R}\left(
x_{2}\right) \right) ,
\end{eqnarray*}
we have 
\begin{equation*}
\me \left( 1-h_{R}\right) \leq \me_{1}\left( 1-g_{R}\right)
+\me_{2}\left( 1-g_{R}\right), 
\end{equation*}
and the claim follows from 
\begin{eqnarray*}
\me \left( \infty \right)  &=& 1-\lim_{R\rightarrow \infty }\me \left(
h_{R}\right) \leq \lim_{R\rightarrow \infty }\me _{1}\left(
1-g_{R}\right) +\lim_{R\rightarrow \infty }\me _{2}\left(
1-g_{R}\right)  \\
&=&\me _{1}\left( \infty \right) +\me _{2}\left( \infty \right) =0.
\end{eqnarray*}
\end{proof}

For a positive Borel measure $\me_{0}$ on $\Omega\times \Omega$, we denote
by
$\left( \me_{0} \right)_{i}$, $i=1,2$, the two measures on $\Omega$
which are margins of $\me_{0}$.

\begin{proposition}\label{m0i}
 Let $\me $ be a mean on $\Omega\times \Omega$. If $\me \left( \infty
\right) =0$, then $\left( \me _{0}\right) _{i}=\left( \me _{i}\right)
_{0}$ for $i=1,2$.
\end{proposition}

\begin{proof}
Let $f \in C_c\left( \Omega\right)$. By Proposition~\ref{minfty} we have 
\begin{equation*}
\me _{0}\left(\widetilde{f}_{i}\right) = \me
\left(\widetilde{f}_{i}\right).
\end{equation*}
Using this equality and the definitions of $(\me_0)_i$ and $(\me_i)_0$ we
get
\begin{equation*}
\left( \me _{0}\right) _{i}\left( f \right) = \me _{0}\left(
\widetilde{f}_{i}\right)=\me \left(
\widetilde{f}_{i}\right) = \me _{i}\left( f \right) =\left( \me
_{i}\right)
_{0}\left( f \right) .
\end{equation*}
\end{proof}

\begin{definition}\label{ovm}
An \emph{operator valued mean} on $\Omega$ is a linear mapping 
\begin{equation*}
M:BC\left(\Omega\right) \longrightarrow \mathcal{L}\left(
\mathcal{H}\right)
\end{equation*}
such that:
\begin{itemize}
\item[(i)] $M\left( f\right) \geq O$ if $f\in BC^{+}\left( \Omega\right)
$;
\item[(ii)] $M\left(1\right) =I$.
\end{itemize}
For an operator valued mean $M$ on $\Omega$ we denote 
\begin{equation*}
M\left( \infty \right) =I-{\rm LUB}\,\left\{ M\left( f\right) \mid
f \in C_{c}^{+}\left( \Omega\right) ,\ f \leq 1\right\} .
\end{equation*}
\end{definition}
\noindent
The least upper bound in the above definition exists by virtue of
Proposition~1 in \cite{NST}. 

Let $M$ be an operator valued mean on $\Omega$. For each $f\in
BC^{r}\left(\Omega\right) $, we have 
\begin{equation*}
M\left( f-\left\| f\right\| _{\infty }1\right) \leq O,\quad M\left(
f+\left\| f\right\| _{\infty }1\right) \geq O.
\end{equation*}
It follows that 
\begin{equation*}
\left\| M\left( f\right) \right\| \leq \left\| f\right\|
_{\infty }.
\end{equation*}
By Theorem~19 in \cite{NST}, there exists a unique positive
operator measure $M_{0}$ on $\Omega$ such that 
\begin{equation*}
M\left( f \right) =\int_{\Omega}f \left( x\right) d M_{0}\left(
x\right) \quad \forall f \in C_{c}\left( \Omega\right), 
\end{equation*}
where the integral is understood in the weak sense. Similarly to the
scalar case we have 
\begin{equation}\label{M0}
M_{0}\left( \Omega\right) =I-M\left( \infty \right) \leq I,
\end{equation}
and, for any $f\in BC\left( \Omega\right)$ we define 
\begin{equation*}
M_{0}\left( f\right) :=\int_{\Omega}f\left( x\right) d M_{0}\left(
x\right).
\end{equation*}

Given an operator valued mean $M$ on $\Omega$ and a unit vector $\psi\in
\mathcal{H}$, we set 
\begin{equation*}
\me_{\psi}\left( f\right) :=\ip{\psi}{M\left( f\right)\psi} \quad
\forall f\in BC\left( \Omega\right) .
\end{equation*}
It is clear that $\me_{\psi}$ is a mean on $\Omega$. By Proposition~1 in
\cite{NST}, 
\begin{equation*}
\me_{\psi}\left( \infty \right) =\ip{\psi}{M\left( \infty \right) \psi}.
\end{equation*}

\begin{proposition}\label{Minfty}
If $M\left( \infty \right)=O$, then 
\begin{equation*}
M\left( f\right) =M_{0}\left( f\right) \quad \forall f\in BC\left(
\Omega\right).
\end{equation*}
\end{proposition}

\begin{proof}
For a unit vector $\psi\in \mathcal{H}$ and a function $f \in
C_{c}\left( \Omega\right) $, we have by definitions 
\begin{equation*}
\left( \me_{\psi}\right) _{0}\left( f \right) =\ip{\psi}{M_{0}\left(f
  \right) \psi},  
\end{equation*}
 and this equality is valid also for any $f\in BC\left( \Omega\right)$.
Since 
\begin{equation*}
\me_{\psi}\left( \infty \right) =\ip{\psi}{ M\left( \infty
\right)\psi} = 0,
\end{equation*}
it follows from Proposition \ref{minfty} that the functional
$\me_{\psi}$ on $ BC\left( \Omega\right)$ coincides with integration with
respect to the measure $\left( \me_{\psi}\right) _{0}$. If $f\in
BC\left( \Omega\right) $, we then have  
\begin{equation*}
\ip{\psi}{M_{0}\left( f\right) \psi} =\left(
\me_{\psi}\right)_{0}\left( f\right) =\me_{\psi}\left( f\right) =\ip{
\psi}{ M\left( f\right) \psi} ,
\end{equation*}
and the claim follows.
\end{proof}

The margins $M_1$ and $M_2$ of an operator valued mean $M$ on
$\Omega\times\Omega$ are defined in an analogous way as in the case of
scalar means.

\begin{proposition}\label{MM}
Let $M$ be an operator valued mean on $\Omega\times \Omega$.
\begin{itemize}
\item[(i)] If $M_{1}\left( \infty \right) =M_{2}\left( \infty \right)
  =O$, then $M\left(\infty \right) =O$;
\item[(ii)] If $M\left(\infty \right) =O$, then $\left( M_{0}\right)
_{i}=\left( M_{i}\right)_{0}$. 
\end{itemize}
\end{proposition}
\begin{proof}
(i) Let $\psi\in\mathcal{H}$ be a unit vector. We have, by definitions,
$\left( m_\psi \right)_i (f)=\ip{\psi}{M_i (f) \psi}$ $\forall f \in
BC(\Omega)$ and $\left( m_\psi \right)_i (\infty)=
\ip{\psi}{M_i (\infty) \psi}$. It follows from Proposition
\ref{mminfty} that $m_\psi(\infty)=0$. Since this is true for any unit
vector, $M(\infty)=O$. \\
(ii) As in the scalar case, by Proposition \ref{Minfty}, we have:
\begin{equation*}
\left(M_0\right)_i\left(f \right)= M_0\left( \tilde{f}_i\right)= M\left(
\tilde{f}_i\right)= M_i\left( f \right)= \left( M_i\right)_0\left( f
\right)
\end{equation*}

\end{proof}

With these results we are ready to prove Proposition \ref{fcoex1}.  

\vspace{0.3cm}

\begin{proof}[Proof of Proposition~\ref{fcoex1}]
Given a function $f:\mathbb{R}\times \mathbb{R}\longrightarrow
\mathbb{C}$ and $\left( q,p\right) \in \mathbb{R}\times \mathbb{R}$,
we denote by $f^{\left(q,p\right) }$ the translate of $f$, 
\begin{equation*}
f^{\left( q,p\right) }\left( x,y\right) := f\left( x+q,y+p\right) \quad 
\forall x,y\in \mathbb{R}.
\end{equation*}
Since $\mathbb{R}\times \mathbb{R}$ (with addition) is
an Abelian group, there exists a mean $\me$ on $\R\times\R$ such that 
\begin{equation*}
\me\left( f^{\left( q,p\right) }\right) =\me \left( f\right) 
\end{equation*}
for all $f\in BC\left( \mathbb{R}\times \mathbb{R}\right)$ and
$(q,p)\in\R\times\R$, (see \cite{AHAI}, Theorem~IV.17.5). 

Let $M_{0}$ be a joint observable of $E_{\rho}$ and $F_{\nu}$. For
each $f\in BC\left(\mathbb{R}\times \mathbb{R}\right) $, for all
$\varphi,\psi\in\hi$ and $q,p\in\mathbb{R}$ we define  
\begin{equation*}
\Theta\left[ f;\varphi,\psi\right] \left( q,p\right) := \ip{W\left(
q,p\right)^{\ast }\varphi}{M_{0}\left( f^{\left(
q,p\right) }\right) W\left( q,p\right)^{\ast}\psi} .
\end{equation*}
Since 
\begin{equation*}
\left\| M_{0}\left( f^{\left( q,p\right) }\right) \right\| \leq
\left\| f^{\left( q,p\right) }\right\| _{\infty }=\left\| f\right\|
_{\infty
}
\end{equation*}
and $W\left( q,p\right)$ is a unitary operator, we have 
\begin{equation*}
\left| \Theta\left[ f;\varphi,\psi\right] \left( q,p\right) \right| \leq
\left\| f\right\| 
_{\infty }\left\| \varphi\right\|\left\| \psi\right\|
\end{equation*}
and hence, $\Theta\left[ f;\varphi,\psi\right]$ is a bounded function.
We claim that $\Theta\left[ f;\varphi,\psi\right] $ is continuous. Since
\begin{equation*}
\Theta\left[ f;\varphi,\psi\right] \left( x+q,y+p\right) =\Theta\left[
  f^{(q,p)};W\left( q,p\right)^{\ast }\varphi,W\left( q,p\right)^{\ast
  }\psi\right] \left(x,y\right) ,
\end{equation*}
it is sufficient to check continuity at $\left( 0,0\right) $. We have
\begin{eqnarray*}
&& \left| \Theta\left[ f;\varphi,\psi\right] \left( q,p\right)
-\Theta\left[
  f;\varphi,\psi \right] \left(0,0\right) \right| \\
&& \qquad \leq \left| \ip{W\left( q,p\right)^{\ast }\varphi}{
  M_{0}\left( f^{\left(q,p\right) }\right) \left(
  W\left( q,p\right)^{\ast}\psi - \psi \right)} \right| \\
&& \qquad \quad +\left| \ip{ \left( W\left(q,p\right)^{\ast}\varphi -
  \varphi \right)}{M_{0}\left(
  f^{\left( q,p\right) }\right)\psi} \right| \\
&&\qquad  \quad +\left| \ip{ \varphi}{ M_{0}\left( f^{\left( q,p\right)}
  -f \right)\psi} \right| \\
&& \qquad \leq \left\| f\right\|_{\infty} \left(
  \left\| \varphi\right\| \left\| W\left(
  q,p\right) ^{\ast}\psi-\psi\right\|
  + \left\| W\left(q,p\right) ^{\ast}\varphi-\varphi\right\|
   \left\| \psi\right\| \right) \\
&& \qquad \quad +\left| \ip{ \varphi}{ 
  M_{0}\left( f^{\left( q,p\right)} -f \right)\psi} \right| 
.
\end{eqnarray*}

As $\left( q,p\right) \rightarrow \left( 0,0\right) $, the first two terms
go to $0$ by the strong continuity of $W$, and the third by the dominated
convergence theorem. We have thus shown that $\Theta\left[
  f;\varphi,\psi\right] \in BC\left( \mathbb{R}\times \mathbb{R}\right)$.

For each $f\in BC\left( \mathbb{R}\times \mathbb{R}\right)$ we can then
define a bounded linear operator $M^{av}\left( f\right) $ by 
\begin{equation*}
\ip{ \varphi}{ M^{av}\left( f\right) \psi} :=\me
\left( \Theta\left[f;\varphi,\psi\right] \right) .
\end{equation*}
It is also immediately verified that the correspondence $M^{av}:BC\left( 
\mathbb{R}\times \mathbb{R}\right) \longrightarrow \lh$ is an operator
valued mean on $\mathbb{R}\times \mathbb{R}$, and a short calculation
shows that 
\begin{equation}
M^{av}\left( f^{\left( q,p\right) }\right) =W\left( q,p\right) ^{\ast
}M^{av}\left( f\right) W\left( q,p\right) . \label{due}
\end{equation}
If $f\in BC\left( \mathbb{R}\right)$ and $(q,p)\in\R\times\R$, we have 
\begin{eqnarray*}
\Theta\left[ \widetilde{f}_{1};\varphi,\psi\right] \left( q,p\right)
&=& \ip{  W\left(q,p\right)^{\ast }\varphi}{ M_{0}\left(
\widetilde{f}_1^{\left(q,p\right) }\right) W\left( q,p\right)
^{\ast}\psi} \\
&=& \ip{  W\left(q,p\right)^{\ast }\varphi}{
W\left(q,p\right)^{\ast } E_{\rho}\left( f\right)W\left( q,p\right)
W\left( q,p\right)^{\ast}\psi} \\
&=& \ip{ \varphi}{ E_{\rho}\left( f\right)\psi} .
\end{eqnarray*}
(In particular, $\Theta\left[ \widetilde{f}_{1};\varphi,\psi\right]$ is a
constant function). Similarly,
\begin{equation*}
\Theta\left[ \widetilde{f}_{2};\varphi,\psi\right] \left(
q,p\right)=\ip{ \varphi}{F_{\nu}\left( f\right)
\psi} .
\end{equation*}
It follows that 
\begin{eqnarray*}
M_{1}^{av}\left( f\right) &=& E_{\rho}\left( f\right),\\
M_{2}^{av}\left( f\right) &=& F_{\nu}\left( f\right).
\end{eqnarray*}
Since $E_{\rho}(\R)=F_{\nu}(\R)=I$, (\ref{M0}) shows that  
\begin{equation*}
M_{1}^{av}\left( \infty \right)=M_{2}^{av}\left( \infty \right) = O.
\end{equation*}
This together with Proposition~\ref{MM} implies that
$M_{0}^{av}\left( \mathbb{R}\times \mathbb{R}\right) =I$ and
\begin{eqnarray*}
\left(M_{0}^{av}\right)_{1} &=& E_{\rho}, \\
\left(M_{0}^{av}\right)_{2} &=& F_{\nu}.
\end{eqnarray*}
By (\ref{due}) the observable $M_{0}^{av}$ satisfies the covariance
condition (\ref{W}).   
\end{proof}

\section*{References}


\end{document}